\documentclass[preprint,twocolumn,5p,super,compress,colorlinks=true]{elsarticle}

\journal{Nature Astronomy}

\usepackage{lineno,hyperref,siunitx,mhchem,booktabs,cuted,titlesec}
\modulolinenumbers[1]

\usepackage[labelfont=bf,justification=justified,singlelinecheck=false]{caption}

\titleformat{\subsection}
  {\small\bfseries}
  {\thesubsection}
  {1em}
  {}

\titleformat{\subsubsection}
  {\small\itshape}
  {\thesubsubsection}
  {1em}
  {}

\makeatletter
\g@addto@macro\@floatboxreset\centering
\renewcommand\@biblabel[1]{#1.}
\makeatother

\newcommand{\DOI}[1]{%
 \newline\noindent\small%
 \textbf{\small{DOI:} }%
 \href{https://dx.doi.org/#1%
 }{#1%
 }%
}
\newcommand{\figref}[1]{\figurename~\ref{fig:#1}}

\begin{document}

\begin{frontmatter}
\title{The seasonal cycle of Titan's detached haze}

\author[JPL,GSMA]{Robert A. West\corref{correspondingauthor}}\ead{robert.a.west@jpl.nasa.gov}
\author[GSMA]{Beno\^{i}t Seignovert}
\author[GSMA]{Pascal Rannou}
\author[JPL]{Philip Dumont}
\author[Hopkins]{~\\Elizabeth P. Turtle}
\author[LPL]{Jason Perry}
\author[JPL]{Mou Roy}
\author[JPL]{Aida Ovanessian}

\address[JPL]{Jet Propulsion Laboratory, California Institute of Technology, Pasadena, CA 91109, USA}
\address[GSMA]{GSMA, Universit\'{e} de Reims Champagne-Ardenne, UMR 7331-GSMA, 51687 Reims, France}
\address[Hopkins]{Johns Hopkins University Applied Physics Laboratory, Laurel, MD 20723, USA}
\address[LPL]{Lunar and Planetary Laboratory, University of Arizona, Tucson, AZ}

\cortext[correspondingauthor]{Corresponding author}

\begin{abstract}
Titan’s \emph{detached} haze, seen in Voyager images in 1980 and 1981 and monitored by the Cassini Imaging Science Subsystem (ISS), during the period 2004-2017, provides a measure of seasonal activity in Titan’s mesosphere with observations over almost one half of Saturn’s seasonal cycle. Here we report on retrieved haze extinction profiles that reveal a depleted layer that visually manifests as a thin layer detached from the main haze below. Our new measurements show the disappearance of the feature in 2012 and its reappearance in 2016, as well as details after the reappearance. These observations highlight the dynamical nature of the detached haze. The reappearance appears congruent but more complex than previously described by climate models. It occurs in two steps, first as haze reappearing at \SI{450(20)}{km} and one year later at \SI{510(20)}{km}. These new observations provide additional tight and valuable constraints about the underlying mechanisms, especially for Titan’s mesosphere, that control Titan’s haze cycle.\\
\DOI{10.1038/s41550-018-0434-z}
\end{abstract}

\end{frontmatter}


Soon after its initial discovery in Voyager images, Titan's photochemical haze exhibited features thought to be directly linked to seasonal behavior \cite{Smith1981,Smith1982,Sromovsky1981}: a hemispheric asymmetry in the main haze, a thin secondary ‘detached’ haze above the main haze and a north (winter) polar hood. The detached haze was observed in the mesosphere (between 300 and 350 km altitude, depending on latitude)\cite{Rages1983}. It was recognized as a feature that could provide clues to the nature of the circulation of Titan’s massive atmosphere.

Using a 1-D model, \citet{Toon1992} argued that the gap separating the detached haze layer from the lower main haze layer could be caused by rising motions which are fast enough to suspend the haze particles. Later, \citet{Rannou2002} employed a General Circulation Model (2D-IPSL-GCM) to study the feedbacks between haze, dynamics and radiation. That model called for upwelling of aerosols in the main haze layer at high latitudes in summer with a horizontal transport toward the winter pole where the returning downwelling removed the haze. The detached haze is interpreted as a zone where aerosols accumulate because their settling speed is balanced by the upward wind speed. 

Alternatively, \citet{Chassefiere1995} proposed that two aerosol production layers, at different altitudes, may explain a global scale gap between the two haze layers. \citet{Lavvas2009} also proposed a model whereby particle microphysical processes alone could explain the nearly-global nature of the detached haze. Cassini observed the detached haze at an altitude of \SI{500}{km}, \SI{150}{km} higher than seen by Voyager\cite{Porco2005} and homogeneous in latitude south of the north polar hood\cite{Seignovert2017}. Initially, the altitude difference posed conceptual problems for the steady-state models described above.

An important new piece of information was provided by \citet{West2011} who showed that the detached haze dropped from \SI{500}{km} in 2005 to \SI{380}{km} in 2010. This work revealed that Voyager had indeed observed the detached haze at a transition period. Monitoring the seasonal cycle of the detached haze layer should provide a way to discriminate between the different possible origins of its formation. \citet{West2011} also report that the 2D-IPSL-GCM predicted a disappearance of the haze near equinox as the overturning circulation stalled prior to reversing. More recent 3D-GCMs\cite{Lebonnois2012,Larson2015} also predicted that, under dominant dynamical process, the detached haze should emerge from the background haze around 2015, at high altitude. 

\section*{Data}

\begin{figure*}[!ht]
 \includegraphics[width=.95\textwidth]{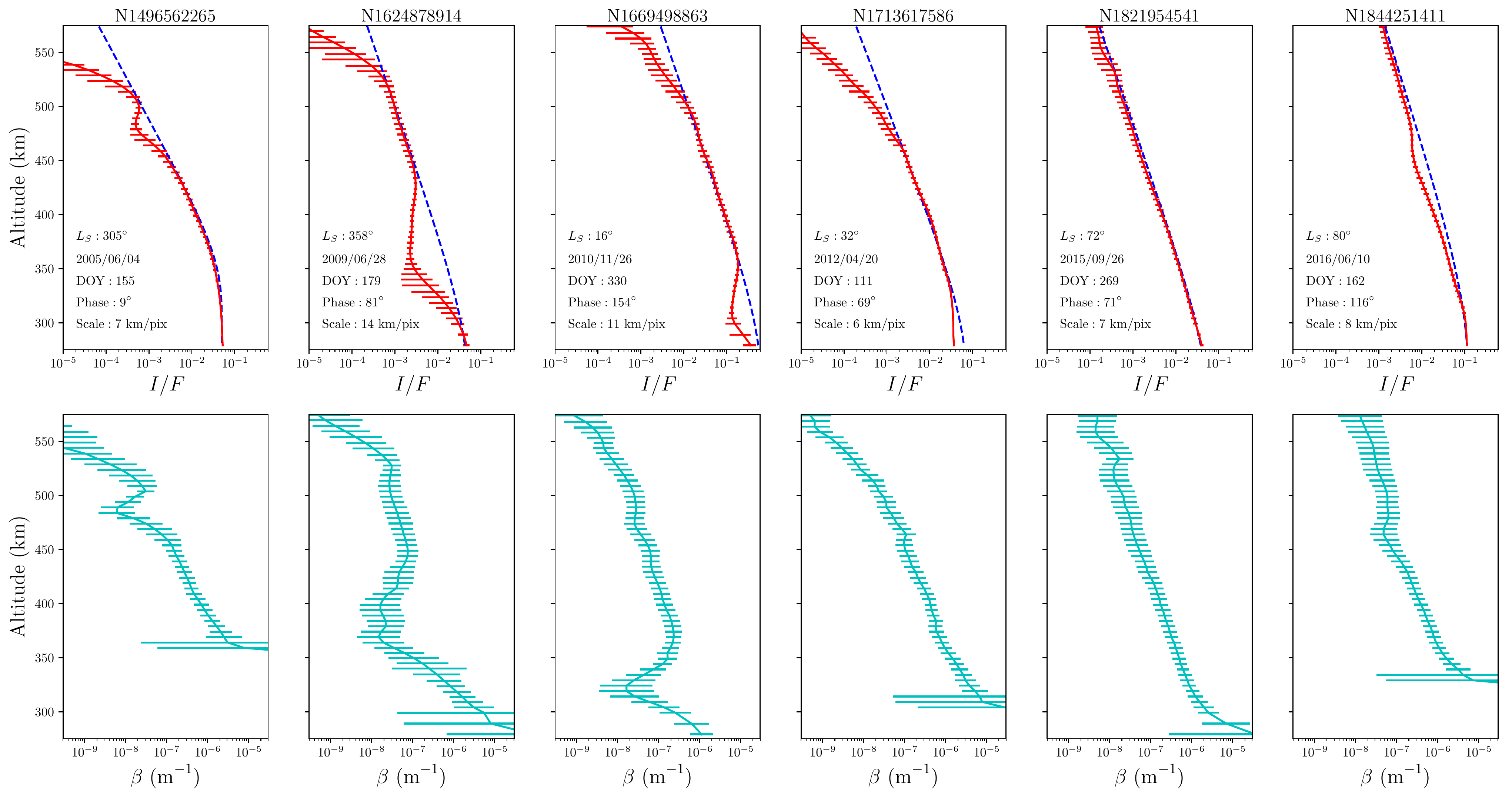}
 \caption{\textbf{Snapshots of Titan limb radiance and retrieved haze extinction.} Upper panel: Vertical profile of the $I/F$ (red) at the equator for six dates; before the equinox and the collapse (2005), during the collapse (2009, 2010, 2011), after the equinox, at the end of the period without the detached haze layer (2015), and finally during the reappearance of the detached haze layer (2016). The uncertainty in $I/F$ is about \num{e-4}. The order of magnitude of $I/F$ differs from panel to panel due to the phase angle differences. The blue dashed lines show a simple model profile with a single scale height. Note that the scale heights differ from panel to panel. This illustrates the existence of a depleted zone which delineates the detached haze. Lower panel: Vertical profiles of the haze extinction corresponding to the $I/F$ shown in the upper panel. The error-bars correspond to the retrieval model uncertainties. We restrict our retrieval to the altitude range between \SI{250}{km} and \SI{700}{km}, and we stop the retrieval as soon as the line-of-sight optical thickness is too high ($\tau_\mathrm{los} > 3$)}
 \label{fig:profiles}
\end{figure*}

To rapidly assess the existence of a weak haze structure we compared $I/F$ limb profiles ($I$ is the intensity and $\pi F$ is the incident solar flux) with a calculated profile for an aerosol diminishing exponentially with fixed sale height\cite{West2011}. Visually there is no sign of a detached haze in near-UV Cassini images after about mid-2012 to early 2016. But between 400 and \SI{480}{km} altitude, the data show a deficit of intensity relative to a haze model in 2016 Day of Year (DOY) 162 (\figref{profiles}).

We analyzed images taken with Cassini’s Imaging Science Subsystem (ISS) Narrow Angle Camera through the CL1-UV3 filter combination (effective wavelength \SI{343}{nm}). The observations were taken over a range of dates and at a variety of phase angles and sub-spacecraft latitudes. For retrievals, we selected $I/F$ vertical profiles of the dayside limb and averaged in a latitude interval between \ang{10}S and \ang{10}N. The pixel scale depends on the distance of the spacecraft to Titan. In our observations, it ranges from about 6 to \SI{12}{km}. To reduce high-frequency oscillations in the retrieval each $I/F$ profile is smoothed with a Gaussian function of \SI{5}{km} full-width at half-maximum and sampled in vertical bins of \SI{5}{km} thickness. $I/F$ values below about \num{e-3} become increasingly uncertain (systematic but image-dependent uncertainty \num{\sim e-4}) due to difficulty determining the background which is dominated by internally scattered light in the camera. 

To retrieve haze extinction coefficient profiles from the $I/F$ profiles, we calculated single-scatter $I/F$ in a spherical-shell geometry\cite{Rages1983,Rannou1997} from \SI{700}{km} altitude to \SI{250}{km}, although the retrieval rarely got as deep as \SI{250}{km}. Layer thickness was \SI{5}{km} in the model. We account approximately for multiple scattering with a plane-parallel radiative transfer model using atmosphere properties defined to fit nadir observations. Then, we integrate the source function at each level of the atmosphere in order to evaluate the ratio of total to single scattering. This value is obtained for the plane of the limb, as a function of the geometry of the observation. It is found to lie between 1 and 1.15 depending on the geometry of the observation. Our method for calculating the multiple scattering ratio was checked using a rigorous spherical-shell model\cite{West2011}. The detail of the model is given in the Methods Detail Section of the Supplementary Material.

\begin{figure*}[!ht]
 \includegraphics[width=.95\textwidth]{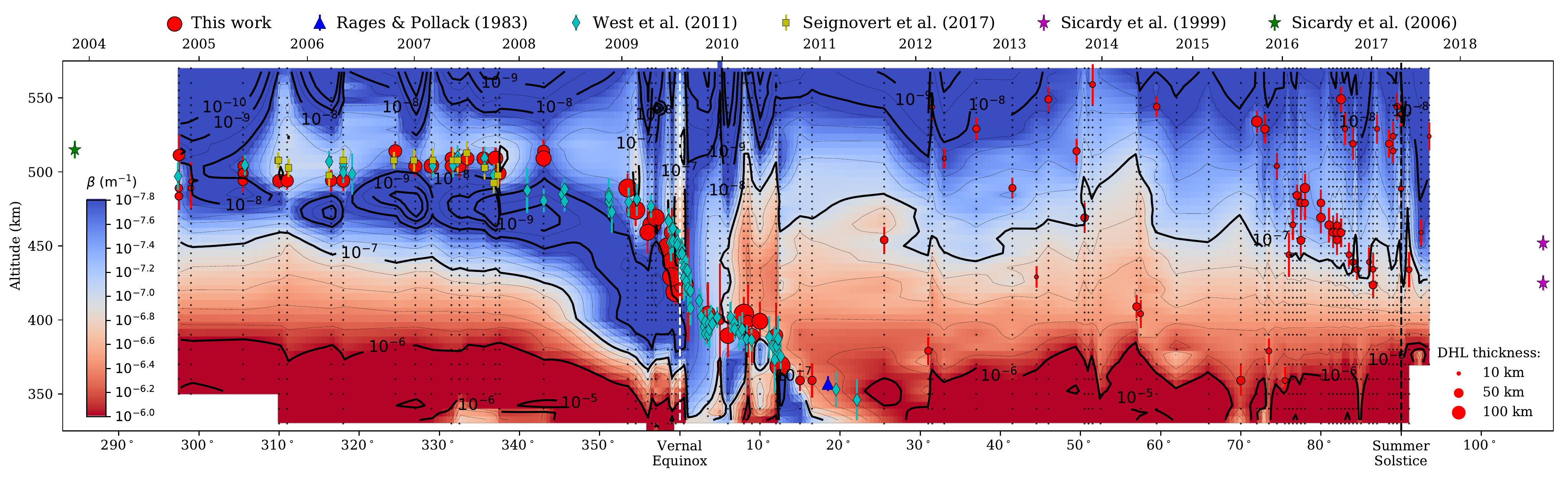}
 \caption{\textbf{Map of the equatorial haze extinction.} Locations of the detached haze detected from $I/F$ profiles as functions of time and altitude\cite{Rages1983,Seignovert2017,West2011}. The size of the red circles (\emph{this work}) are proportional to the vertical thickness of the detached haze layer. The local maximum of $I/F$ at the limb provides a clear reference to assign the detached haze altitude prior to mid-2012.
 After 2012 and especially in 2016 and 2017 the haze deficit is weak, and a local maximum is not apparent. Instead we identify the altitude of the detached haze with the minimum in the second derivative of $I/F$ with respect to altitude.
 Note that we do not plot points if there was no detached haze feature in the image. Note also that the single measurement by \citet{Rages1983} one Titan year earlier than the Cassini measurement falls at the same altitude (\SI{\sim 360}{km}) as the Cassini measurement.
 Vertical dotted lines indicate locations of images we incorporated into the contour map, including those that do not show a detached haze. The stellar occultation of 1988\cite{Sicardy1999} detected two peaks of temperature gradient possibly linked to detached hazes, but was not able to probe above \SI{500}{km} while the stellar occultation of 2003\cite{Sicardy2006} probed up to \SI{600}{km} and could detect the temperature inversion correlated to the DHL.
 The error-bars correspond to one pixel scale on the images.}
 \label{fig:map}
\end{figure*}

We used data taken at different times, at different phase angles and possibly with a haze layer experiencing natural variability at timescales smaller than Titan’s \num{\sim 7}-year seasons to produce a time-altitude extinction map. \figref{map} includes results from \citet{West2011} and new measurements using an automated technique for determining the altitude of the detached haze. For the new measurements, we chose the best sample of the 317 UV3 images available to get the highest limb, phase and temporal coverage. The images are taken with at least one day separation. Our sampling is not evenly distributed due to orbital constrains and mission schedule but at least \SI{90}{\percent} of our samples are separated by less than 120 days. The two main gaps of data are encountered between 28 March 2008 and 25 January 2009 (302 days) and 26 November 2010 and 9 September 2011 (286 days). There are no UV3 images during those periods.

The properties of aerosols may change with time or the haze may fluctuate due to gravity wave activity, for example. Some observations taken a few hours or days apart demonstrate that the haze fluctuates at very short time scales. Moreover, the navigation (including the altitude scale) of observations may be uncertain by few km to more than \SI{10}{km} (about one pixel), depending on the resolution and the nature of the image. 
Undoubtedly some of these issues may degrade the quality of the retrieval and the clarity of the haze extinction evolution at the equator as a function of time and altitude. To partially compensate for these uncertainties and bias, we smoothed the map by aligning the extinction profiles at \SI{425}{km} to follow a smooth curve passing through the actual retrieved values at this level, and then scaling the vertical profiles. The resulting map of extinction for the equatorial region, not totally devoid of artifacts, is given in \figref{map}, along with the locations of the detached haze determined by the maxima of the $I/F$ profiles.

\figref{map} shows a stable detached haze layer prior to 2009 and a drop over the equinox as reported by \citet{West2011}. We determine here that the collapse starts with a rapid fall, at the mean free fall speed, of \SI{80}{km} during 100 days, followed by a slower fall before its disappearance \SI{3}{years} after the equinox. During the slower fall, the detached haze comes to the same location as observed by Voyager exactly one Titan year before, suggesting a very regular seasonal cycle. Until the detached haze feature faded, merging with the main haze below \SI{300}{km} in 2012 Day of Year (DOY) 272, (3.14 years after equinox), it could be readily seen in UV and Blue filter Cassini ISS images with pixel scale better than \SI{10}{km\per pixel} (Figure 5 of ref.\cite{West2016}).

\begin{figure*}[!ht]
 \includegraphics[width=.75\textwidth]{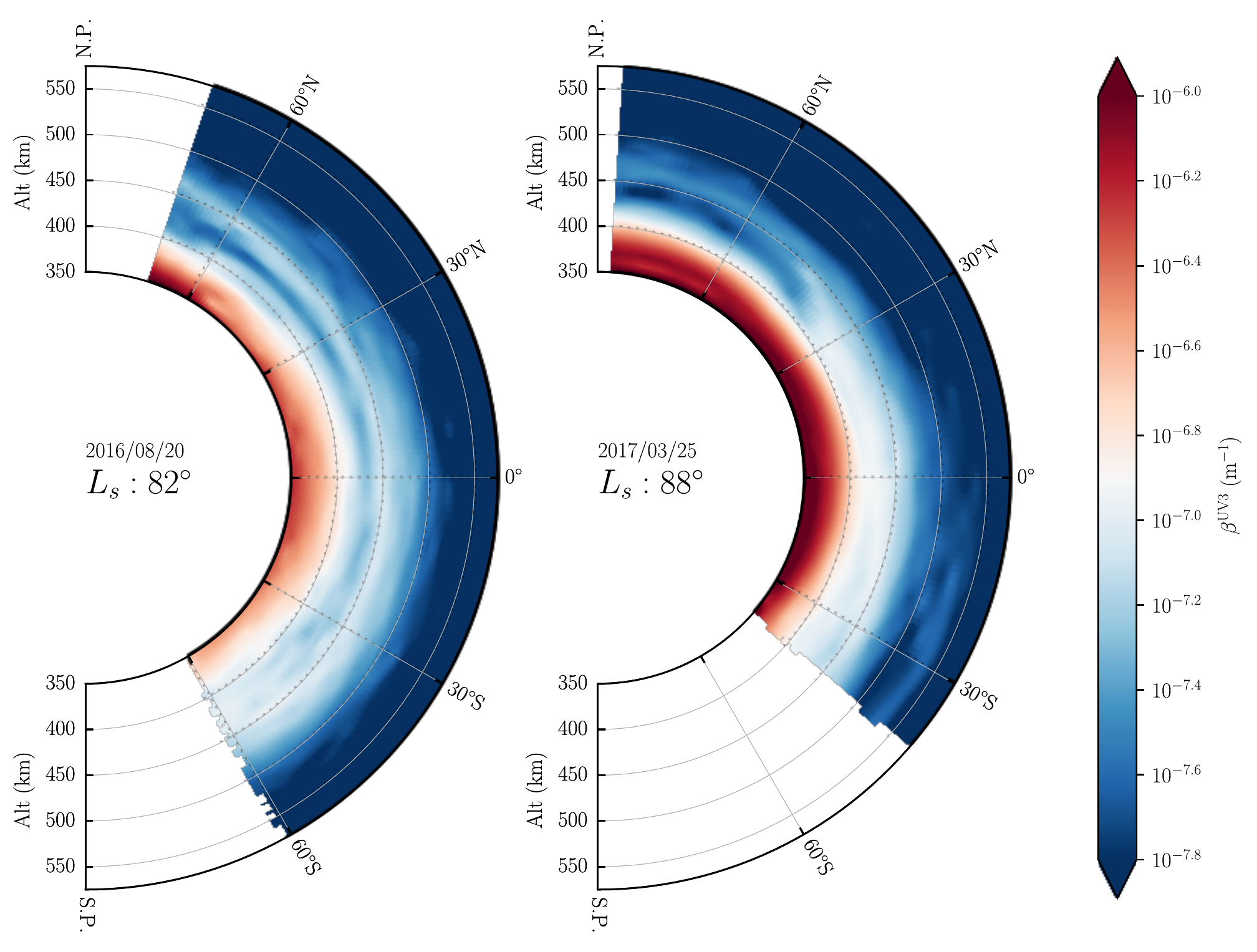}
 \caption{\textbf{The detached haze layer at the northern summer solstice.} Haze extinction as functions of latitude and altitude on two occasions during the reappearance of the DHL. The detached haze appeared in early 2016 and became a permanent structure at about \SI{480}{km} altitude, slowly descending during one terrestrial year down to \SI{450}{km} before vanishing in early 2017. The upper panel shows this haze about in the middle of this time period. Then, a new detached haze appeared around \SI{500}{km} in 2017 while the first detached layer can still be detected in the northern polar region (lower panel). This second layer is almost complete but not as well formed as the first one in 2016.}
 \label{fig:limb}
\end{figure*}

In the period between 2012 and the beginning of 2016, the haze does not present major structures, excepted very localized and temporary oscillations in the haze vertical profile, sometimes with local maxima, possibly related to wave activity. These structures are not stable in time and can be found at any altitude and latitude. From DOY 21 of 2016, several profiles suddenly showed a marked inflexion in the $I/F$ profiles, and an inversion in extinction, which appears frequently and then permanently at about \SI{480}{km}. Indeed, the detached haze layer visually emerged from the background, in UV photometry \SI{6.5}{years} after the equinox, within an extremely short time interval (\SI{\sim 2.5}{months}). This first detached layer descended by 80 km in about one year and is now vanishing (as of May, 2017, it still persists in the northern polar region). From DOY 263 of 2016, a new detached haze appeared at \SI{510}{km} and strengthened during 2017. The corresponding extinction profiles show a persistent local maximum that can be clearly associated to the emergence of a detached haze layer. We verified that these new features also have large-scale latitudinal extent, as expected for the detached haze layer. Both detached hazes, when developed, cover a large part of the planet (\figref{limb}). As for the detached haze before 2009, this haze appears separated from the main haze due to diminished extinction at some altitudes suggesting that exactly the same processes are at work, although with much weaker contrasts. 

Recent data from the Cassini Infrared Radiometer/Spectrometer (CIRS) also show a post-equinox depleted layer forming in early 2016 at \SI{400}{km}, in the northern hemisphere\cite{Vinatier2016}. This altitude is \SI{20}{km} lower than the extinction depletion zone retrieved from UV ISS images, within error bars of CIRS and ISS. Such independent results confirm the recent major change occurring in the mesospheric haze layer. Future comparisons between UV and IR data will provide constraints on aerosol properties during this rapid change.

Climate models\cite{Rannou2002,Lebonnois2012,Larson2015,Rannou2004} find that during most of the Titan year, the detached haze results in the accumulation of submicron aerosols conveyed by a broad subsolar (summer) ascending branch of the stratospheric circulation. Aerosols accumulate at the altitude level where their settling speed compensates the upward vertical component of the wind. There, they are subsequently driven toward the winter pole by meridional winds producing a second scattering layer above the main haze. Both models and observations show that the detached haze is also composed of macromolecules formed at higher levels\cite{Larson2015,Rannou2004,Cours2011,Lavvas2011}. The details of the detached haze in models depend upon its physical properties (fractal dimension, monomer radius, number of monomers, which influence sedimentation velocity) and how they factor into the coupling with the meridional circulation.

\citet{Lavvas2009} emphasized the possible causal role of the transition from small monomers to larger aggregates, and how this transition may be related to the detached haze. We have some evidence that both monomers and aggregates populate the detached layer\cite{Cours2011}. In the Cassini ISS images the intensities drop into the background at altitudes above about \SI{550}{km} where we expect the aerosol to be dominated by monomers. Particle optical cross sections are larger at smaller wavelengths, so further investigation at shorter UV wavelengths from the Cassini Ultraviolet Imaging Spectrograph (UVIS) can yield new insight\cite{Koskinen2011,Lavvas2016}.

At equinox, climate models predict a reversal of the stratospheric seasonal circulation and a collapse of the detached haze similar to the observed behavior in 2009\cite{West2011}. Models suggest that the vertical circulation, which was able to suspend aerosols at around \SI{500}{km} altitude before the equinox, disappeared and no longer supported the detached haze layer. Particle free fall near equinox implies that, during about \SI{100}{days}, ascending winds became negligible compared to the sedimentation speed. After a temporary stall of six month in 2010, the subsequent slower fall also appears to be a free fall. According to models, the detached haze should reappear at high altitude several years after the equinox. 

The free-fall, disappearance, and reappearance observed in our data grossly resemble predictions from models. However, some noticeable differences remain between models and observations. Our retrievals show, as \citet{West2011} described, that the visual appearance of the detached haze is due to an aerosol depletion layer between the main and the detached layers and not a supplementary haze added to the main haze above a given altitude, as suggested by models\cite{Rannou2002,Lebonnois2012,Larson2015}. There is a small phase difference and a significant altitude difference between models and observations\cite{West2011}. The altitude difference is due to an unrealistic upper limit to the GCM dictated by computing resources. The 3D-GCMs appear in much better agreement for the timing, with a DHL starting to fall at equinox ($L_s \approx \ang{0}$) and reappearing several months before $L_s = \ang{90}$ (May 2017). \citet{Larson2015} predicted an earlier return of the detached haze than was observed at $L_s = \ang{70}$ (November 2015). However, their plot exhibits the aerosol mass mixing ratio which cannot be directly compared to optical properties. \citet{Lebonnois2012} predict the return of the DHL a bit after $L_s = \ang{90}$. Beyond the timing, the GCMs do not capture the complexity of the reappearance in a two-step process as shown by observations.

\section*{Implications}

Differences between models and observations may have several causes. In models, the haze production function may be too simple; aerosol geometrical structure which controls the mechanical behavior may be too simple or erroneous. Model setups, primarily the spatial resolution of the models (\SI{10}{km} vertically and \SIrange{150}{200}{km} horizontally) may be too coarse regarding the structure of the detached haze layer. Finally, it should be noted that GCM results are time-averaged over several terrestrial months. This erases short term or faint structures and time variability, as we find for the detached haze layer after 2015.

Several implications come from this work. First, in the mesosphere, the response time of the haze to the dynamics is very short (\SI{e6}{s})\cite{Rannou2002} and the feedback of the haze onto the global circulation involves a much longer timescale and deeper layers. Therefore, the detached haze does not directly modify the seasonal cycle and it can be considered as a direct tracer of the mesospheric circulation. Events reported here give a direct and valuable view of the timing of the circulation turnover at equinox and bring tight constraints on the underlying dynamical processes. Moreover, as noted previously, the detached haze appears correlated to a local thermal signature with a strong positive gradient below the detached haze layer\cite{Porco2005,Sicardy2006,Fulchignoni2005}. Such a link is also obvious in coupled GCMs\cite{Lebonnois2012,Larson2015,Rannou2004} and may produce a local feedback between the circulation and the detached haze itself\cite{Cours2011}. This would make it possible to monitor the DHL seasonal cycle after the Cassini era, through its thermal signature, with future stellar occultations as it was monitored in the past\cite{Sicardy1999,Sicardy2006}. Secondly, higher-resolution images (e.g. Fig. 10 of ref.\cite{Porco2005}), not investigated here, suggest that fine structure within the detached haze fluctuates on timescales of hours or days. The latitudinal structure of the detached haze will also be a subject of future study. \figref{limb} also shows a marked spatial variability. The microphysical timescale to form aerosols\cite{Cabane1993} as large as monomers is about \SI{e6}{s} and larger aerosols coming from the main haze are even older. Thus, we conclude that only small-scale dynamical processes can induce such rapid changes. A close inspection of this layer (using latitude, altitude and time variation), would give a new insight on rapid and small-scale processes at work in this layer.

Further progress on small-scale variations can move forward by investigating fine details of higher-resolution images where local horizontal and vertical variations of intensity can be seen (e.g. Fig. 10 of ref.\cite{Porco2005}). In rare cases, multiple images of the same part of the atmosphere were taken several minutes or hours apart. Such image sets can reveal how the haze evolves at short timescales, possibly capturing wave activity. Additionally, modeling gravity waves could address small-scale variations, while further improvements to the GCM could address the larger-scale variations we report.

We show here the behavior of Titan’s equatorial haze in the upper stratosphere/lower mesosphere on a seasonally relevant time scale. A deficit in haze density over a narrow altitude range produces the visual appearance of a detached layer which descends with time, sometimes (near equinox) at the rate expected for particle sedimentation. After equinox the haze disappeared for a few years, then reappeared sporadically and weakly at equatorial latitudes. These details will be valuable as observables feeding dynamical and haze microphysical models which at this time are able to reproduce some but not all of the detail we see.



\renewcommand\thefigure{S\arabic{figure}}    
\setcounter{figure}{0}    

\section*{Methods (Supplementary Material)}
\subsection*{Retrieval method}
\subsubsection*{Scattering model}
To retrieve the haze extinction coefficient profiles from the $I/F$ profiles, we first model $I/F$ with a single-scattering model in spherical-shell geometry\cite{Rages1983,Rannou1997} from above \SI{700}{km} altitude to \SI{250}{km} altitude.  We checked the calculations with a spherical-shell RT calculator that accurately calculates both single and multiple scattering11 but that model is too slow for retrievals.  We grid the atmosphere in $n$ layers and, assuming single scattering by a spherical atmosphere, we write the outgoing radiance factor $I/F_n$, for the bottom of the layer $n$, as following:

\begin{equation}\label{eq:rt_model}\scriptsize
    I/F_n = \sum\limits_{i=1}^{2n} \int\limits_{x_{i-1}}^{x_i}
    \frac{\varpi_j P_j\left(\Theta_s\right)}{4}
    \exp\left( -\tau^0_i - \tau^1_i \right)
    \beta\left(z\left(x\right)\right)
    \varrho_{ms}\left(z\left(x\right)\right) \mathrm{d}x
\end{equation}

where the summation is performed on the $2n$ segments which are defined by the intersections of the line of sight and the spherical shells defining the layer boundaries.  The impact factor $z_0$, that is the lowest altitude reached by the line of sight, is given by the bottom of the $n^\mathrm{th}$ layer and consequently $I/F_n = I/F(z_0)$. Therefore, each layer of the atmosphere is crossed twice. $x$ is the abscissa along the line of sight. $\tau^0_i$ and $\tau^1_i$ are the opacities along the incident and emergent path. $P_j\left(\Theta_s\right)$ and $\varpi_j$ are respectively the phase function at the scattering angle of the observation $\Theta_s$ and the average single scattering for the layer $j$, where the index $j = i$ if $j < n + 1$ and $j = 2n + 1 - i$ if $j > n$. $\beta\left(z\left(x\right)\right)$ is the extinction at the latitude $z$, and $\varrho_{ms}\left(z\left(x\right)\right)$ the multiple scattering ratio. Here, the altitude $z\left(x\right)$ is the local altitude at point of abscissa $x$ along the line of sight, and is defined as $z\left(x\right) = \sqrt{z_0^2 + x^2}$. 

We account for the multiple scattering as a correction factor computed separately $\varrho_{ms}\left(z\left(x\right)\right)$. In the detached layer, the multiple scattering is produced by the light coming from the atmosphere below. So, we assume a model which matches nadir observations in UV.  Then, with this model, we integrate the source function coming from the Titan disk below, at each level of the atmosphere and we compare to the direct solar flux in order to evaluate the ratio of multiple scattering to single scattering as a function of the altitude. This value is obtained for the plane of the limb, as a function of the geometry of the observation. This ratio is found between 1 to 1.15 depending on the geometry of the observation. Our evaluation of the multiple scattering ratio was computed using a more sophisticated model\cite{West2011}.

\subsubsection*{Retrieval technique}

With this model, the $I/F(z)$ profile only depends on the extinction profile $\beta(z)$ of the haze and the illumination, viewing and phase angles. We assume that there is no horizontal inhomogeneity of the atmospheric properties along the line of sight. Both the phase function and single scattering albedo are computed with a model of scattering by fractal aggregates. We used aggregates of \SI{1000}{spherical} grains of \SI{50}{nm} which produces a phase function that matches, to first order, the $I/F$ as a function of phase angle at \SI{400}{km} and \SI{500}{km} altitude. The single-scattering albedo is set to \num{0.9}.

We need to retrieve a set of $N$ extinction values $\beta_i$ (composing the vector $\beta$) that matches the $n$ values of the radiance factor, $I/F_i$ (composing the vector $I/F$).  In the discrete form of Equation~\eqref{eq:rt_model}, the scattered intensity at a given level $i$, $I/F_i$, is a function of the extinction $\beta_j$ in the levels $j$ where $j \le i$\cite{Rages1983}.

The $I/F_i$ as a function of extinction $\beta_i$ is of the form of a nonlinear triangular system. Due to the nonlinearity, we have no simple analytical way to solve it (it is not a matrix system). But there are several other ways to solve it -- typically an onion-peeling method where the $\beta_i$ are solved sequentially from the top to the bottom. We rather decided to use a Levenberg-Marquardt retrieval where we retrieve the vector $\beta(\beta_1, \beta_2,..., \beta_N)$ that best fits the intensity vector $I/F( I/F_1, I/F_2,..., I/F_N)$. A Levenberg-Marquardt retrieval is useful because, in principle, the system can be solved exactly and the solution is unique. We restrict our retrieval to the altitude range between \SI{250}{km} and \SI{700}{km}, and we stop the 
retrieval as soon as the line-of-sight optical thickness $\tau_\mathrm{los} > 3$.

\subsubsection*{Choice of the phase functions}

\begin{figure}[!ht]
 \includegraphics[width=\linewidth]{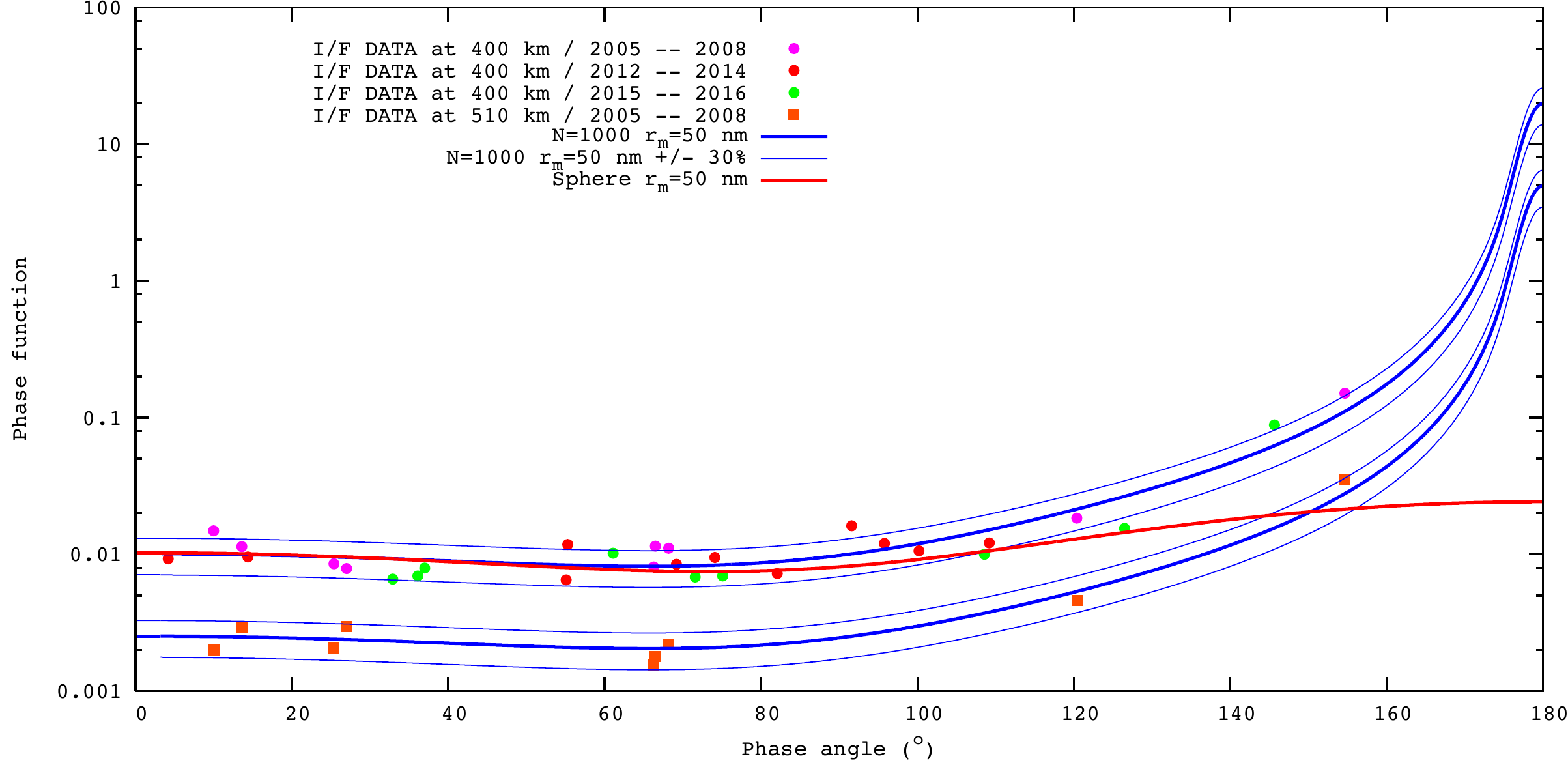}
 \caption{$I/F$ data collected at \SI{400}{km}, in the main haze, and at \SI{510}{km}, in the detached haze, as functions of the phase angle during selected time intervals for which the hazes appear stable. The thick blue curves show the phase function of aggregates of \SI{1000}{monomers} of \SI{50}{nm} radius, scaled to fit the data. The thin curves are the same phase function increased or decreased by \SI{30}{\percent}. The red line is the phase function of a single monomer.}
 \label{fig:phase_function}
\end{figure}

\figref{phase_function} shows observed $I/F$ as a function of the phase angle, collected in the main haze layer (at \SI{400}{km}) during the periods 2005-2008 and 2012-2016, and at the level of the detached haze (\SI{510}{km}) during the period 2005-2008. These periods were chosen because the hazes were visually stable and, thus, the variation of $I/F$ as a function of the phase angle should approximately reflect the phase function. This neglects time variability of the haze extinction (which produces the scattering of the data around the model curves) and the effect of multiple scattering. We find that all phase functions for fractal aggregates larger than several hundred monomers of \SI{50}{nm} radius match these curves. Beyond this threshold, there is no more sensitivity of the phase function at the observed phase angles (that is in the range \ang{0} to \ang{160}). From these data, we retrieved phase functions assuming aggregates of \SI{1000}{ monomers} for all levels. We also plot, for comparison, the phase function for a sphere of \SI{50}{nm} radius, representative of a single monomer.

\subsection*{Free-fall calculation}

\begin{figure*}[!hb]
 \includegraphics[width=.7\linewidth]{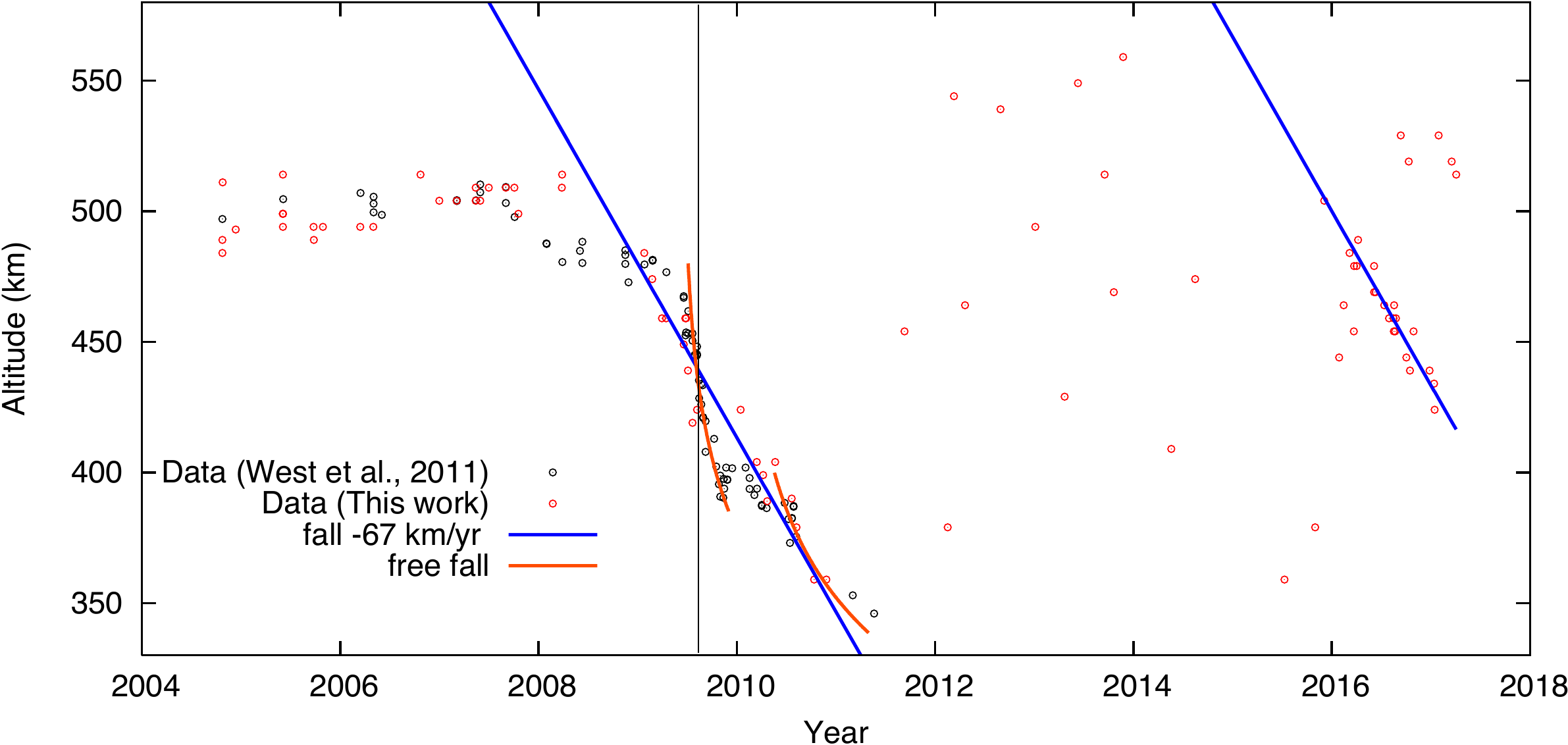}
 \caption{Locations of the detached haze layer as a function of time. The thin vertical line shows the time of the equinox. The path of a particle under free fall (under gravitational and atmosphere drag forces) is shown with the orange curve. The blue lines show a fall at the constant rate of \SI{-67}{km\per year}. For the free fall, we used the sedimentation speed for a fractal aggregate of dimension $D_f = 2$ made of \SI{50}{nm} radius monomers, published by \citet{Larson2015} to calculate analytically this trajectory. Under a free fall, such a particle drops by \SI{80}{km} in \SI{100}{days}. This trajectory is comparable to the trajectory of the detached haze during its collapse after equinox (\figref{map}).}
 \label{fig:free_fall}
\end{figure*}

To estimate the path of a free fall motion from \SI{480}{km}, we consider the settle speed of a spherical particle with a radius of \SI{50}{nm} in the free molecular regime\cite{Fuchs1989}. The settle speed follows a law that we model as $v(z) \approx v_0 \exp(a \cdot z)$\cite{Larson2014} where $a = \partial \log(v)/\partial z$ and $v_0$ the extrapolated settle speed at $z = 0$. The algebraic integration of $\mathrm{d}x = v(t) \mathrm{d}t$ yields the altitude of the particle under free fall from an altitude $z_i$ at $t = 0$ as the function $z(t) = z_i - 1/a \cdot \log(1 + c \cdot t)$ where  $c = a \cdot v_0 \cdot \exp(a \cdot z_i)$. On Titan, we evaluate (from Figure 6 of ref.\cite{Larson2014}) these numerical values as $a \approx \SI{2.2e-5}{\per\meter}$, $v_0  \approx \SI{7.1e-7}{\meter\per\second}$. \figref{free_fall} gives the path of free fall from $z_i$ = \SI{480}{km} as a function of time obtained with these equations. It can be compared to the drop of the detached haze layer after equinox in \figref{map}. 

\vfill\null

\begin{figure*}[!hb]
    \includegraphics[width=.7\linewidth]{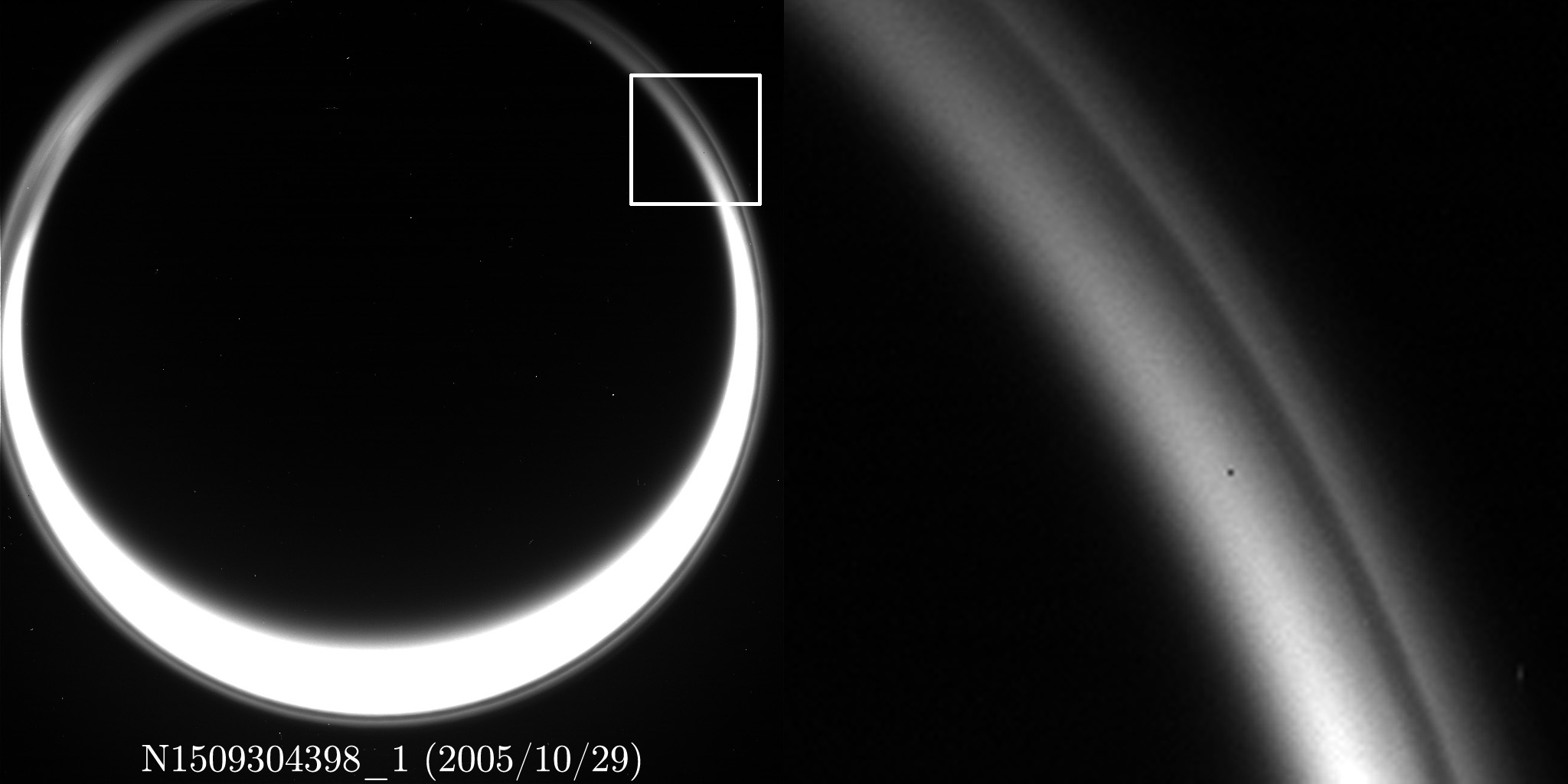}
    \includegraphics[width=.7\linewidth]{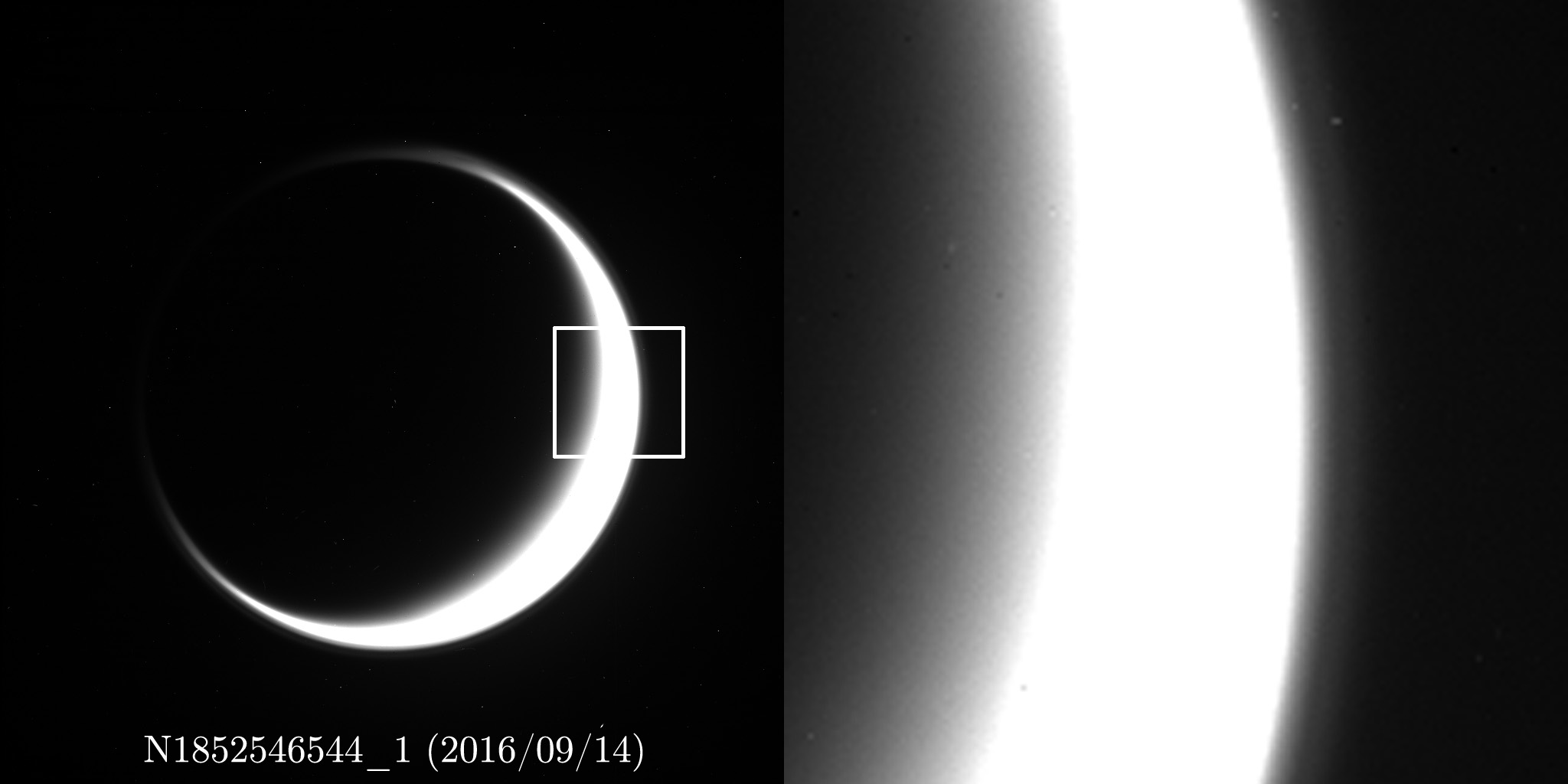}
    \caption{High phase angle images of the main and detached hazes of Titan before the collapse of the detached haze (top) and at the time of the reappearance (bottom). The observations are quite similar in terms of geometry with \ang{155} and \ang{142} phase angle for the 2005 and 2016 images. Images on the right half of the figure show the equatorial region corresponding to the squares in the left half.}
    \label{fig:imgs}
\end{figure*}

\clearpage

\subsection*{Detached haze: Appearance in images}

The detached haze layer is, over a long period of time, a continuous feature along the limb of Titan, above the main haze layer.  The scattered intensity is linked to the integrated opacity along the line of sight. This is especially true at the detached haze level where haze is optically thin. Therefore, our results concerning the haze extinction directly reflects the observed intensities. We display here two images of Titan (\figref{imgs}) with the detached haze layer before the equinoctial collapse (29 Oct. 2005) and during the reappearance (14 Sep. 2017). In both case, the detached haze is a thin shell that almost completely surrounds the main haze layer (except for the winter polar vortex poleward of \ang{55} latitude. However, the intensity ratio between the detached and main haze layer is clearly not the same at the two dates. The detached haze appears much fainter in 2017 than in 2005, demonstrating that the newly formed detached layer is optically much weaker that the detached layer before the collapse.

\section*{Data Availability}
The data used in this study are available from the \href{https://pds.nasa.gov/}{Planetary Data System} Imaging Node.  Calibration software for the images is also available from the PDS.  Further inquiries about the findings of this study are available from the corresponding author upon reasonable request.

\section*{Acknowledgment}
Research was supported by the Cassini-Huygens mission, a cooperative endeavour of NASA, ESA and ASI managed by JPL/Caltech under a contract with NASA. Part of this work was performed by the Jet Propulsion Laboratory, California Institute of Technology. Part of this work was done while R.A.W. was hosted at the Universit\'{e} de Reims Champagne-Ardenne with the support of the R\'{e}gion Champagne-Ardenne, and simultaneously with the support of the JPL Senior Research Scientist Leave programme. P.R. thanks the Agence Nationale de la Recherche (ANR Project \emph{APOSTIC} no.~11BS56002, France).

\section*{Author contributions}
R.A.W. contributed original data and measurements. B.S. contributed measurements and figures. P.R. contributed aerosol retrievals and discussion of haze and climate models. P.D. assisted with scattering models. E.P.T., J.P. and M.R. assisted with data acquisition. A.O. assisted with measurements.

\section*{Competing interests}
The authors declare no competing interests.

\bibliography{Biblio}

\end{document}